\documentclass{PoS}

\usepackage{cite}
\usepackage{graphicx,epsfig,amsmath,amssymb}

\newcommand{\mhh}{m_{hh}}

\newcommand{\pth}{p_{T,h}}
\newcommand{\ct}{c_t}
\newcommand{\ctt}{c_{tt}}
\newcommand{\chhh}{c_{hhh}}
\newcommand{\cg}{c_{ggh}}
\newcommand{\cgg}{c_{gghh}}
\def\nn{\nonumber}

\title{Shape analysis in Higgs boson pair production}

\ShortTitle{Shape analysis in Higgs boson pair production}

\author{Matteo Capozi, \speaker{Gudrun Heinrich}\\
        Max Planck Institute for Physics, F\"ohringer Ring 6, 80805
        Munich, Germany\\
        E-mail: \email{mcapozi@mpp.mpg.de}, \email{gudrun@mpp.mpg.de}}

\abstract{We
study the impact of  anomalous couplings in the Higgs sector on the shape of the Higgs boson pair invariant mass
distribution at NLO. 
Our analysis is based on a five-dimensional coupling parameter space relevant for
Higgs boson pair production in gluon fusion, in the framework of a
non-linear Effective Field Theory.
In particular, we present a clustering procedure into
certain shape types based on unsupervised machine learning, with the
aim to infer information about the underlying parameter space from a
given shape type.}

\FullConference{14th International Symposium on Radiative Corrections (RADCOR2019)\\ 
9-13 September 2019\\
		Palais des Papes, Avignon, France}

\begin{document}

\section{Introduction}

Higgs boson pair production in gluon fusion is an important process to extract information about the trilinear Higgs boson coupling $\lambda$.
Recent LHC measurements~\cite{Aad:2019uzh,Sirunyan:2018two} have used this process to constrain $\lambda$ to the range $-5.0\leq \lambda/\lambda_{\rm{SM}}\leq 12.0$~\cite{Aad:2019uzh}. Combined constraints, based on indirect measurements from single Higgs production processes and direct limits from double Higgs boson production lead to ~$-2.3 \leq \lambda/\lambda_{\rm{SM}}\leq 10.3$~\cite{ATLAS:2019pbo}. However the latter constraints strongly depend on the  assumption that all
deviations from the SM expectation are stemming from a modification of the trilinear coupling only, while the other couplings are fixed to their SM values.

%%%%%%%%%%%%%%%%%%%%%%%%%%%%%%%%%%%%%%%%%%%%%%%%%%%

On the theory side, various attempts have been made to come up with 
constraints on $\chhh=\lambda/\lambda_{\rm{SM}}$ that are largely model independent~\cite{Falkowski:2019tft,Chang:2019vez,DiLuzio:2017tfn,DiVita:2017eyz}, suggesting  $|\chhh|\lesssim 4$ for a new physics scale in the few TeV range.
Recent phenomenological studies about the precision that could be reached for the self-couplings at the (HL-)LHC and future hadron colliders are summarised in Refs.~\cite{Dawson:2018dcd,Basler:2018dac,Cepeda:2019klc,DiMicco:2019ngk}.

\medskip

Higgs boson pair production in gluon fusion in the SM has been calculated at leading order in Refs.~\cite{Eboli:1987dy,Glover:1987nx,Plehn:1996wb}, and at NLO in Ref.~\cite{Dawson:1998py} in the $m_t\to\infty$ limit (``heavy top limit, HTL''), rescaled with the full Born matrix element. 
%In Ref.~\cite{Maltoni:2014eza} an approximation called ``FT$_{\rm{approx}}$"  was introduced,  which contains the full top quark mass dependence in the real radiation, while the virtual part is calculated in the rescaled heavy top limit.
The NLO QCD corrections with full top quark mass dependence have been calculated in Refs.~\cite{Borowka:2016ehy,Borowka:2016ypz,Baglio:2018lrj}.
Implementations of the full NLO QCD corrections in parton shower Monte Carlo programs are also available~\cite{Heinrich:2017kxx,Jones:2017giv,Heinrich:2019bkc}.
The uncertainties due to the chosen top mass scheme have been assessed in Ref.~\cite{Baglio:2018lrj}. 

In the HTL, the NNLO QCD corrections have been computed in Refs.~\cite{deFlorian:2013uza,deFlorian:2013jea,Grigo:2014jma,Grigo:2015dia,deFlorian:2016uhr}, the N$^3$LO corrections in Ref.~\cite{Chen:2019lzz}.
%These results have been improved in various ways: they have been supplemented by an expansion in $1/m_t^2$ in~\cite{Grigo:2015dia}, and soft gluon resummation has been performed at NNLO+NNLL level in~\cite{deFlorian:2015moa}. 
 The calculation of Ref.~\cite{deFlorian:2016uhr} has been combined with results including the top quark mass dependence as far as available in Ref.~\cite{Grazzini:2018bsd}, and the latter has been supplemented by soft gluon resummation in Ref.~\cite{deFlorian:2018tah}. 
Analytic approximations to the top-mass dependent two-loop amplitudes in several limits, as well as phase space integrals for the NNLO real radiation, have been studied in Refs.~\cite{Grober:2017uho,Bonciani:2018omm,Xu:2018eos,Davies:2018ood,Davies:2019xzc}. Complete analytic results in the high energy limit have been presented in Ref.~\cite{Davies:2018qvx}; the latter have been combined with the full NLO results, in the regions where they are more appropriate, in Ref.~\cite{Davies:2019dfy}.

The effects of operators within an Effective Field Theory (EFT) description of Higgs boson pair production have been studied at NLO in the HTL in Refs.~\cite{Grober:2015cwa,Grober:2016wmf,Maltoni:2016yxb,Grober:2017gut}. EFT studies at NNLO in the HTL are also available~\cite{deFlorian:2017qfk}.
In Ref.~\cite{Buchalla:2018yce} for the first time the full NLO QCD corrections have been combined with an EFT approach to study BSM effects.

\vspace*{3mm}

It is well known that Higgs boson pair production in gluon fusion is a process where delicate cancellations occur between triangle-type contributions, which contain the trilinear Higgs coupling,  and box-type contributions.
A deviation of the trilinear coupling from its SM value would change this interference pattern and thus lead to characteristic shape changes in the di-Higgs invariant mass distribution $\mhh$.
However, the question arises whether such shape changes could as well be induced by a combination of other anomalous couplings in the Higgs sector, thus faking an anomalous value for $\lambda$.
Therefore, in Ref.~\cite{Capozi:2019xsi},  we performed a shape analysis based on a 5-dimensional coupling parameter space and proposed a method to associate certain shapes with distinct regions in the parameter space.

The idea of a shape analysis has been pursued already in various ways based on LO studies, see e.g.  Refs.~\cite{Chen:2014xra, Azatov:2015oxa, Dawson:2015oha,Carvalho:2015ttv,Carvalho:2016rys,Carvalho:2017vnu}.
In Ref.~\cite{Carvalho:2015ttv}, a cluster analysis has been proposed to define 12 benchmark points in a 5-dimensional non-linear EFT parameter space which result from clusters of ``similar" shapes.
 The similarity measure in this case is based on a binned likelihood ratio using LO predictions for the observables $\mhh$, $\cos\theta^*$ and $\pth$.
 In Ref.~\cite{Buchalla:2018yce} it was analysed how the $\mhh$ and $\pth$ distributions change when going from LO to NLO for the benchmark points defined in Ref.~\cite{Carvalho:2015ttv}, and it was found that the NLO corrections can have a significant impact on the shapes.

We do not attempt to define new benchmark points here, but rather would like to identify more global patterns in the $\mhh$ spectrum which can be attributed to coupling configurations.
To this aim we apply an unsupervised learning algorithm to identify patterns in the shapes of the $\mhh$ distribution. 
Then we use the {\tt KMeans} clustering algorithm from {\tt scikit-learn}~\cite{scikit} and ask for a classification of the shapes into a given number of clusters, and finally relate the clusters to the underlying parameter space.

\section{Parametrisation of anomalous couplings in the Higgs sector}

Our studies are based on an effective Lagrangian in a non-linear Effective Field Theory (``Higgs Effective Field Theory, HEFT'')~\cite{Buchalla:2015wfa,Buchalla:2018yce} relevant for 
Higgs boson pair production, assuming CP conservation:
\begin{align}
{\cal L}\supset 
-m_t\left(c_t\frac{h}{v}+c_{tt}\frac{h^2}{v^2}\right)\,\bar{t}\,t -
c_{hhh} \frac{m_h^2}{2v} h^3+\frac{\alpha_s}{8\pi} \left( c_{ggh} \frac{h}{v}+
c_{gghh}\frac{h^2}{v^2}  \right)\, G^a_{\mu \nu} G^{a,\mu \nu}\;.
\label{eq:ewchl}
\end{align}
In the SM $c_t=c_{hhh}=1$ and $c_{tt}=c_{ggh}=c_{gghh}=0$.
The chromomagnetic operator is absent in (\ref{eq:ewchl}) because it 
contributes to $gg\to hh$ only at higher order in the counting
underlying the HEFT Lagrangian~\cite{Buchalla:2018yce}.
The coefficients $c_{ggh}$ and $c_{gghh}$ are related in SMEFT (``SM
Effective Field Theory'')~\cite{DiMicco:2019ngk}, however in HEFT they
are a priori independent.

%For more details about the difference between HEFT and SMEFT we refer to Refs.~\cite{Buchalla:2018yce,DiMicco:2019ngk}.
We produce our data  using the differential distributions calculated
in Ref.~\cite{Buchalla:2018yce}, parametrised in terms of 23 coefficients $A_i$ for each coupling combination occurring in the (differential) NLO cross section, which allows for a fast evaluation:
\begin{align}
\frac{d \sigma}{dm_{hh}} =& A_1c_{t}^4+A_2c_{tt}^2+A_3c_{t}^2c_{hhh}^2+A_4c_{ghh}^2c_{hhh}^2 + A_5c_{gghh}^2 + A_6c_{tt}c_{t}^2+A_7c_{t}^3c_{hhh} \nn\\ &+A_8c_{tt}c_{t}c_{hhh}+A_9c_{tt}c_{ggh}c_{hhh}+A_{10}c_{tt}c_{cgghh}+A_{11}c_{t}^2c_{ggh}c_{hhh}+A_{12}c_{t}^2c_{gghh} \nn\\ & + A_{13}c_{t}c_{hhh}^2c_{ghh}+A_{14}c_{t}c_{hhh}c_{gghh}+A_{15}c_{ggh}c_{hhh}c_{gghh} + A_{16}c_{t}^3c_{ggh}+A_{17}c_{t}c_{tt}c_{ggh} \nn\\ & +A_{18}c_{t}c_{ggh}^2c_{hhh}+A_{19}c_{t}c_{ggh}c_{gghh}+A_{20}c_{t}^2c_{ggh}^2+A_{21}c_{tt}c_{ggh}^2+A_{22}c_{ggh}^3c_{hhh}\nn\\
&+A_{23}c_{ggh}^2c_{gghh} \;.
\label{eq:Ai_mhh}
 \end{align}
The coefficients $A_i$ are available in bins of width 20\,GeV  from
250\,GeV to 1050\,GeV.

\section{Shape classification and clustering}
\label{sec:classification}

%\subsection{Classification and clustering}
%\label{sec:shapes}

There is no need to use machine learning for shape classifications, it
is certainly possible to write an analyser function that checks the
slopes of the distributions and puts them into predefined shape classes. 
However it turned out that this procedure is cumbersome, as it should
take into account bin-by-bin variations as well as more global
features of the $\mhh$ spectrum, and it is hard to extend to a larger
number of shape types.
Therefore, to avoid the bias introduced by the ``manual'' definition of the shape
types, and to find a more flexible classification  which
can be extended easily to more shape types, 
we devised a different approach to the classification problem,
using unsupervised learning techniques.

We constructed a classification of the shapes of the $\mhh$
distribution into a given number of shape clusters, where we did not predefine what the clusters should look like.
For this purpose we used an autoencoder to find common patterns in the data.
The setup is implemented using {\sc Keras}~\cite{keras} and {\sc TensorFlow}~\cite{tensorflow}.
As input data we used 30 bins of width 20\,GeV for the normalised $\mhh$ distributions. 
We trained a neural network based on a set of $10^5$ distributions,
produced by variations of the coupling parameters in ranges as specified below.

To reduce uncertainties, for example due to overfitting, 
we produced ten different autoencoder models,
where for each model we picked $10^4$ random points from the training set for
validation. 
The ten models all have the same goal, 
but are starting from different training and validation sets and thus a different
initialisation of the weights.
We trained the autoencoder for each model over 10000 epochs using {\tt
  Adam}~\cite{Kingma:2014vow} as optimiser and the root mean square
error to define the loss function.
The ten different encoded training data sets are then fed to a
classification algorithm, where we employed the {\tt KMeans}
clustering algorithm from {\tt scikit-learn}~\cite{scikit}, asking for
a classification into a given number of clusters, where we
tested classifications into 4--8 clusters.
\begin{figure}[htb]
\centering
\includegraphics[width=0.8\linewidth]{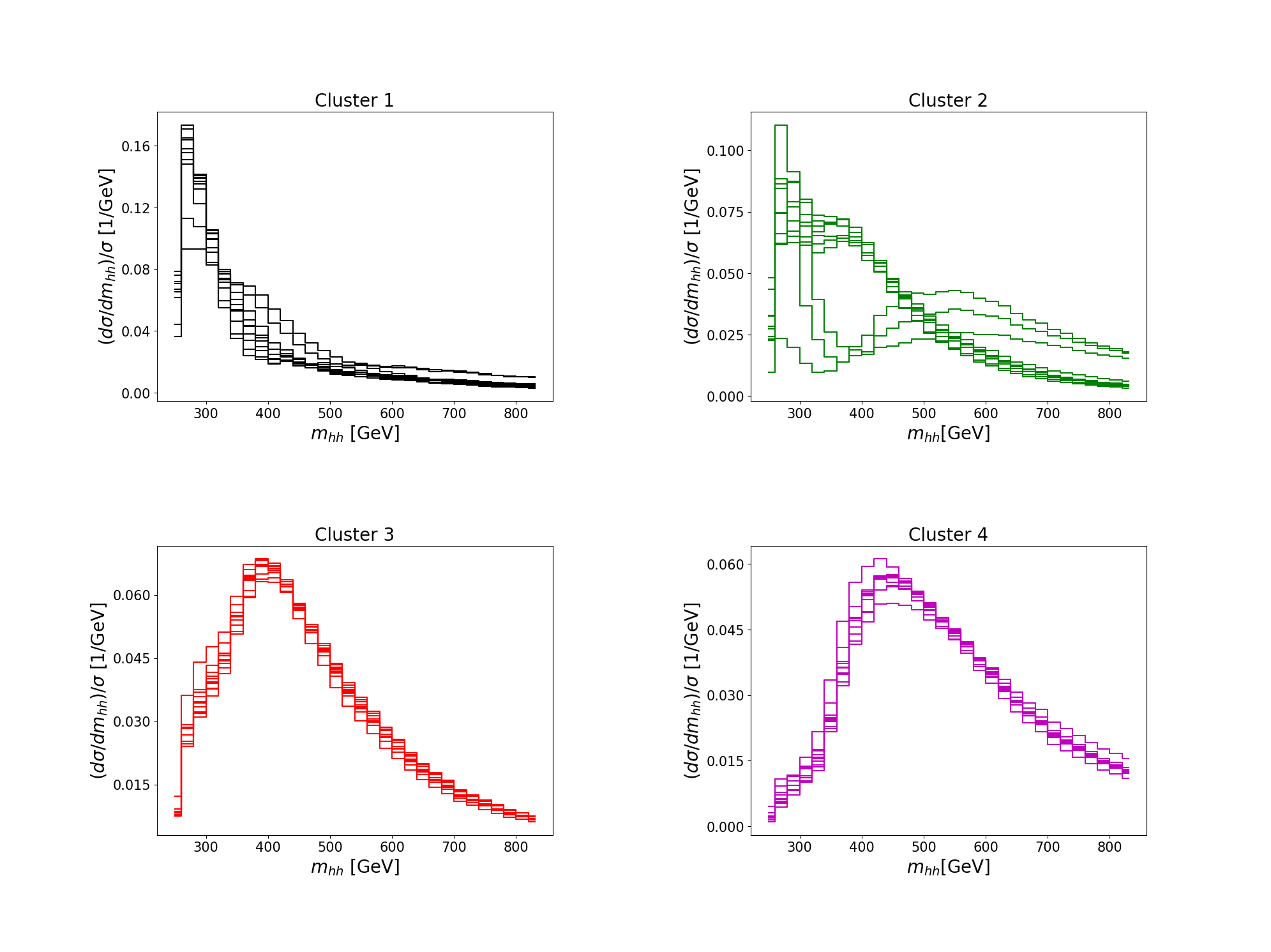}
\includegraphics[width=\linewidth]{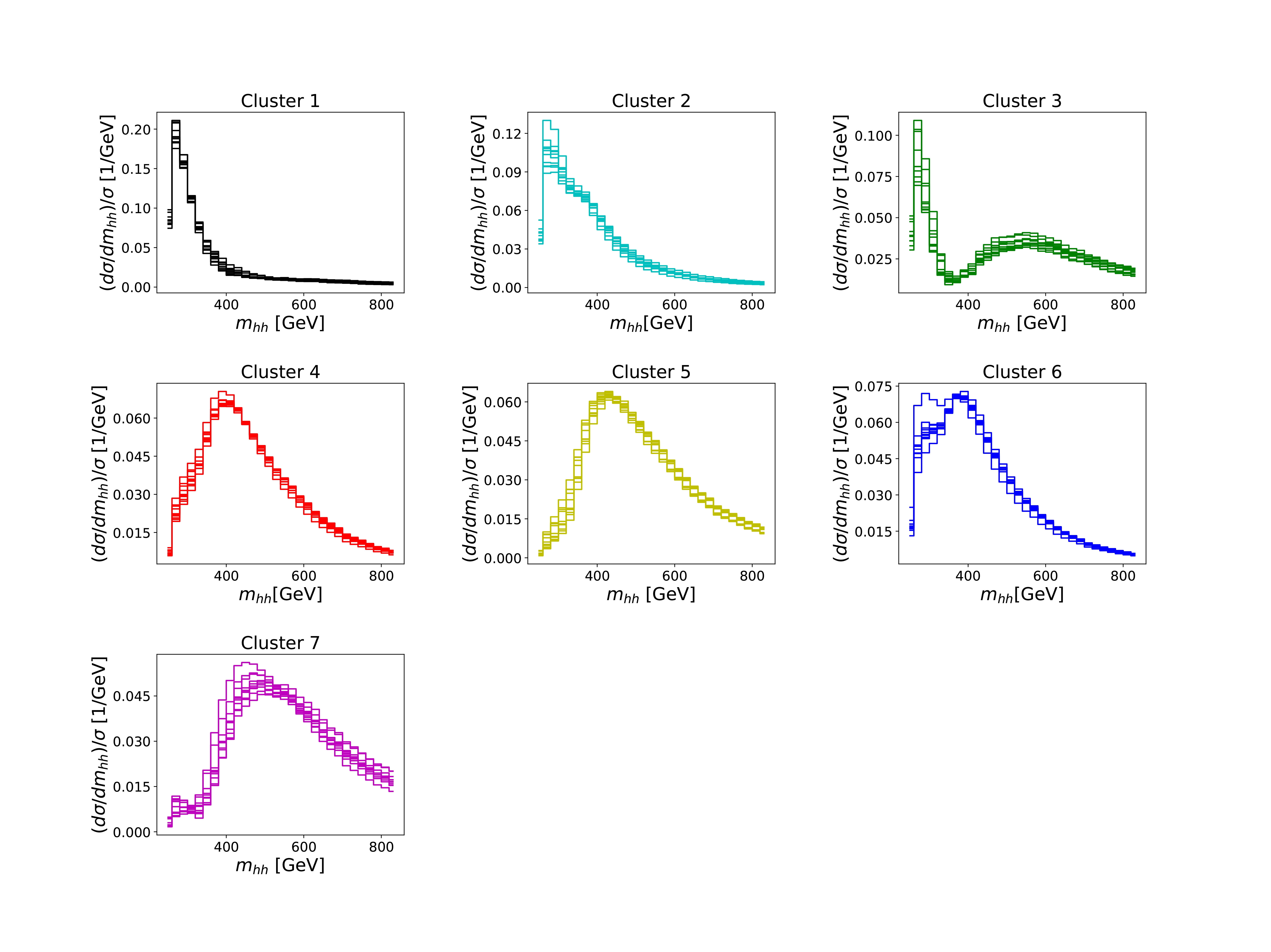}
  \caption{The clusters obtained by asking for a classification into
    four or
    seven shape types. The cluster centres obtained from 10 different
    encoder models are shown, in
    the colour codes used later to visualise the underlying parameter space.}
  \label{fig:KMeans7}
\end{figure}
Asking the {\tt KMeans} algorithm to find 4 and 7 clusters yielded the
shape types shown in Fig.~\ref{fig:KMeans7}.
The curves denote
the cluster centres determined by the  {\tt KMeans} algorithm, for
each of the ten encoder models.

One can see from Fig.~\ref{fig:KMeans7} that in the case of clustering
into four shape types, cluster 2 contains shapes which vary substantially.
In contrast, for seven shape types, the cluster centres obtained
from the ten different encoder models are quite similar.
Applying the procedure to 4--8 clusters revealed that seven clusters seemed to be the optimal number to
capture distinct shape features, while 
defining more than seven clusters did not lead to useful additional features, but rather to the 
tendency to focus on local minima in the clustering space.

To combine the results from the ten autoencoder models,  we adopted
the ``majority vote'' method, i.e. for each of the ten
outcomes, a given point in the coupling
parameter space gets a label (``vote'') corresponding to the cluster it belongs
to. The final cluster assigned to that point is the one which
collected the largest number of votes.

\clearpage

\section{Parameter space underlying the clusters}

In this section we show how the parameter space relates to the clusters if we ask for 4 or 7 clusters. 
Our results for the $gg\to hh$ cross sections at NLO are produced for a centre-of-mass energy of $\sqrt{s}=13$\,TeV,  using PDF4LHC15$\_$nlo$\_$100$\_$pdfas~\cite{Butterworth:2015oua} parton distribution functions interfaced via LHAPDF, along with the corresponding value for $\alpha_s$.
The masses have been set to $m_h=125$\,GeV, $m_t=173$\,GeV and the top quark width has been set to zero.
We study the differential cross section as a function of five anomalous couplings, varying them in the ranges specified below:
\begin{equation}
 \ct \in [0.5,1.5], \; c_{hhh} \in [-3,8], \; c_{tt} \in [-3,3], \;
 c_{ggh},\; c_{gghh} \in [-0.5,0.5]\;.
 \label{eq:ranges}
\end{equation}
The ranges are motivated by current experimental constraints. For
$\chhh$ we use a smaller range than constrained by experiment in order
to focus more on the range where interesting shape changes occur.
In order to visualise the results, we project out 2-dimensional slices of the 5-dimensional parameter space, fixing 
the other three couplings to their SM values.
\begin{table}[htb]
\begin{center}
\begin{tabular}{|c| c | c |}
\hline
 Cluster&type&colour\\
\hline
\hline
\multicolumn{3}{|c|}{4 clusters}\\
\hline
\hline
 1& enhanced low $\mhh$&  black \\
 2& double peak/shoulder&  green \\
 3& SM-like&  red \\
 4& enhanced tail &  magenta \\
\hline
\hline
\multicolumn{3}{|c|}{7 clusters}\\
\hline
\hline
 1& enhanced low $\mhh$ &  black \\
 2& enhanced low $\mhh$, slowly falling or shoulder&  cyan\\
 3& enhanced low $\mhh$, second local maximum above  $\mhh\simeq 2m_t$&  green \\
 4& SM-like &  red \\
 5& SM-like with enhanced tail & yellow \\
 6& close-by double peaks or shoulder left&  blue \\
 7& no steep slope at low $\mhh$, enhanced tail&  magenta \\
\hline
\end{tabular}
\end{center}
\caption{Clusters and shape types with corresponding colour codes for the classification into four and seven clusters.}
\label{tab:clusters_colours}
\end{table}
The colour codes are shown in Fig.~\ref{fig:KMeans7}, and are also listed in Table~\ref{tab:clusters_colours}.

\begin{figure}[htb!]
  \centering
\includegraphics[width=0.8\linewidth]{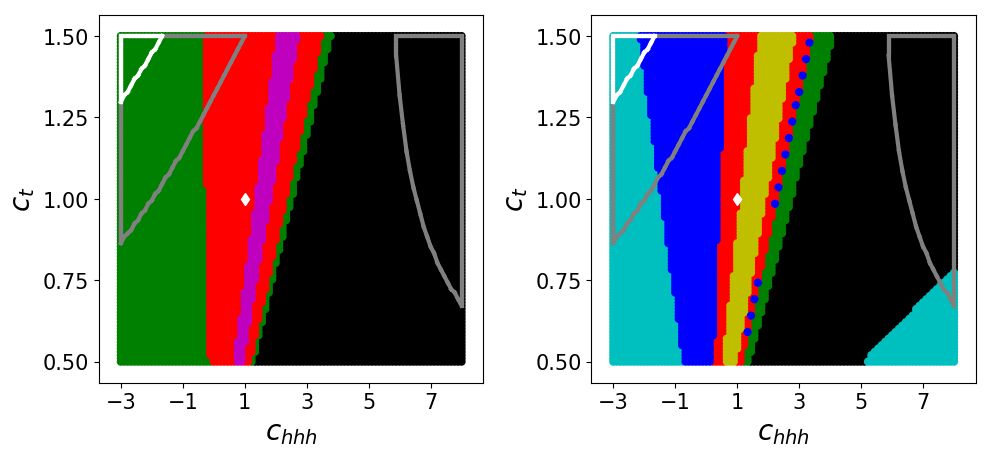}
  \caption{Shape types produced by variations of $\ct$ versus $\chhh$. 
  Left: 4 clusters, right: 7 clusters.
  The areas outside the silver and white curves are regions where
  the total cross section exceeds $6.9\times \sigma_{SM}$ and $22.2 \times \sigma_{SM}$, respectively.
These values are motivated by the current ATLAS/CMS limits at $\sqrt{s}=13$\,TeV~\cite{Aad:2019uzh,Sirunyan:2018two}.
The white diamond denotes SM parameter point.
The colour code is given in Table \ref{tab:clusters_colours}.}
  \label{fig:ct_c3}
\end{figure}
% 
%%%%%%%%%%%%%%%%%%%%%%%%%%%%%%%%%%%%%%%%%%%%%%%%%%%%%
\begin{figure}[htb!]
  \centering
\includegraphics[width=0.8\linewidth]{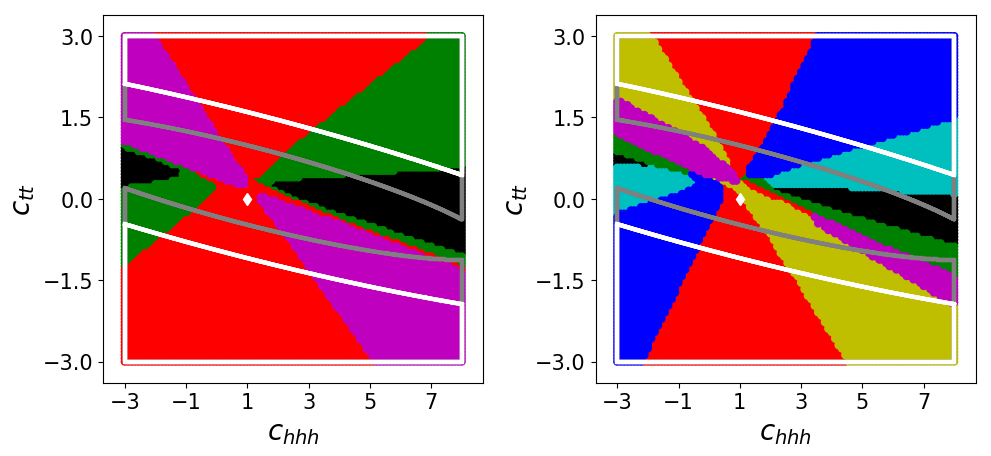}
%  \label{fig:c3_ctt}
%\includegraphics[width=0.8\linewidth]{plots/cggh-cgghh.png}
  \caption{Shape types produced by variations of $\ctt$ versus $\chhh$. Left: 4 clusters, right: 7 clusters. The areas outside the silver and white curves are regions where
  the total cross section exceeds $6.9\times \sigma_{SM}$ and $22.2 \times \sigma_{SM}$, respectively.}
  \label{fig:chhh_ctt}
\end{figure}
In  Fig.~\ref{fig:ct_c3} we display shape types resulting from
variations of $\ct$ versus $\chhh$. We observe that the clustering
into seven clusters provides a much more refined distinction of
SM-like shapes from distributions with an enhanced tail or a doubly
peaked structure than the classification into four clusters.
The figure clearly shows that small variations of $\chhh$ can easily distort the SM-like shape, while the shape is more robust against variations of $\ct$.
Fig.~\ref{fig:chhh_ctt} shows $\ctt$ versus $\chhh$,
where we see that the interplay between $\chhh$ and $\ctt$ can lead to all shape types.
Small deviations of these couplings from the SM value already can have a substantial effect on the shape.
Furthermore, it becomes apparent that $\ctt$ values different from zero enhance the total cross section, such that limits on the total cross section combined with shape information allows to constrain $\ctt$.
%

\begin{figure}[htb]
  \centering
\includegraphics[width=0.5\linewidth]{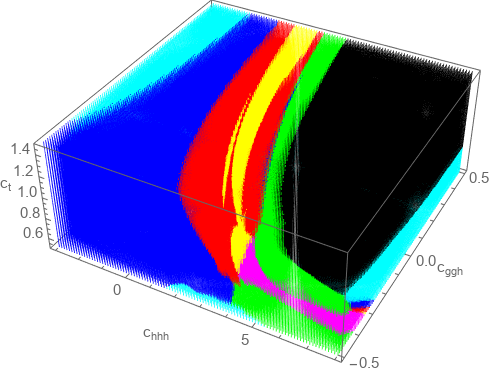}
  \caption{Three-dimensional visualisation of shape types produced
    by variations of $\ct$, $\chhh$ and $\cg$. For the parameters not shown we used $\cgg=0.5\cg$ and $\ctt=0.05\ct$,  simulating the SMEFT
    situation.}
  \label{fig:SMEFT}
\end{figure}

%%%%%%%%%%%%%%%%%%%%%%%%%%%%%%%%
Projections of all coupling parameter combinations onto two-dimensional planes show~\cite{Capozi:2019xsi} that the parameters $\chhh$ and $\ctt$
have the largest influence on the shape.
In SMEFT,  $\ctt$ is suppressed compared to $\ct$ by one order of the large new physics
scale~\cite{DiMicco:2019ngk}.
Furthermore, SMEFT imposes the relation  $\cg=2\cgg$.
Using this relation and imposing that $\ctt$ amounts to 5\% of $\ct$,
we obtain a 3-dimensional parameter space simulating the SMEFT
situation, which is visualised in
Fig.~\ref{fig:SMEFT}.

%\clearpage

\section{Conclusions}

We have identified shape clusters for the Higgs boson pair invariant
mass distribution $\mhh$ based on an unsupervised machine learning approach.
Using these clusters we investigated how anomalous couplings in the
Higgs sector affect the shape of the $\mhh$ distribution.
We found that the trilinear coupling $\chhh$ and an effective $t\bar{t}hh$
coupling, $\ctt$, have the largest influence on the shape, while variations of the top Yukawa
coupling and effective gluon-Higgs couplings in the experimentally
allowed ranges have a smaller shape-changing effect. 
%An largely enhanced low $\mhh$ region is almost impossible to achieve with
%the SM value of $\chhh$.
As the SM-like shape is not very robust against variations of $\chhh$,
combined information about the shape of the $\mhh$
spectrum and the total cross section has some potential to reveal
New Physics effects at the High-Luminosity-LHC.

\section*{Acknowledgements}
We would like to thank Gerhard Buchalla, Alejandro Celis, Long Chen,
Victor Diaz, Stephan Jahn, Stephen P.~Jones, Matthias Kerner, Gionata
Luisoni, Ludovic Scyboz and Johannes Schlenk for collaboration on 
Higgs boson pair production and for useful discussions.
This research was supported in part by the COST Action CA16201
(`Particleface') of the European Union and
by the Deutsche Forschungsgemeinschaft (DFG) under Germany's
Excellence Strategy EXC-2094-390783311.

\bibliographystyle{JHEP}
\bibliography{refsML}
%\begin{thebibliography}{99}

%\end{thebibliography}

\end{document}